\documentclass{article}
\usepackage{amsmath,amsthm,bm,mathrsfs}
\usepackage[normalem]{ulem}	% Part of the standard distribution
\usepackage[T1]{fontenc}
\usepackage{xcolor}
\usepackage{graphicx}
\usepackage{epsfig}
\usepackage[round]{natbib}
 \newtheoremstyle{theorem}{6pt}{6pt}{\rm}{}{\sffamily}{ }{ }{}
 \theoremstyle{theorem}

 \newtheoremstyle{algorithm}{6pt}{6pt}{\rm}{}{\sffamily}{ }{ }{}
 \theoremstyle{algorithm}

 \newtheoremstyle{lemma}{6pt}{6pt}{\rm}{}{\sffamily}{ }{ }{}
 \theoremstyle{lemma}

\newtheoremstyle{case}{6pt}{6pt}{\rm}{}{\sffamily}{. }{ }{}
 \theoremstyle{case}

 \newtheoremstyle{statement}{6pt}{6pt}{\rm}{}{\sffamily}{ }{ }{}
\theoremstyle{statement}

 \newtheoremstyle{corollary}{6pt}{6pt}{\rm}{}{\sffamily}{ }{ }{}
 \theoremstyle{corollary}

  \newtheoremstyle{definition}{6pt}{6pt}{\rm}{}{\sffamily}{ }{ }{}
 \theoremstyle{definition}

\newtheoremstyle{example}{6pt}{6pt}{\rm}{}{\sffamily}{ }{ }{}
\theoremstyle{example}

\newtheoremstyle{remark}{6pt}{6pt}{\rm}{}{\sffamily}{ }{ }{}
\theoremstyle{remark}

\newtheoremstyle{approximation}{6pt}{6pt}{\rm}{}{\sffamily}{ }{ }{}
\theoremstyle{approximation}

\newtheoremstyle{scheme}{6pt}{6pt}{\rm}{}{\sffamily}{ }{ }{}
\theoremstyle{scheme}

\newtheoremstyle{Algorithm}{6pt}{6pt}{\rm}{}{\sffamily}{ }{ }{}
\theoremstyle{Algorithm}

\newtheoremstyle{Assumption}{6pt}{6pt}{\rm}{}{\sffamily}{ }{ }{}
\theoremstyle{Assumption}

\newtheoremstyle{proposition}{6pt}{6pt}{\rm}{}{\sffamily}{ }{ }{}
\theoremstyle{proposition}

\newtheoremstyle{hypo}{6pt}{6pt}{\rm}{}{\sffamily}{ }{ }{}
 \theoremstyle{hypo}

  \newtheoremstyle{Step}{6pt}{6pt}{\rm}{}{}{ }{ }{}
 \theoremstyle{Step}

\newcommand{\be}{\begin{equation}}
\newcommand{\ee}{\end{equation}}
\newcommand{\bea}{\begin{eqnarray}}
\newcommand{\eea}{\end{eqnarray}}

\newcommand{\nd}{\noindent}

\begin{document}

\title{Stochastic Model for Tumor Control Probability: Effects of Cell Cycle and (A)symmetric Proliferation}
\author{{\sc Andrew Dhawan$^{1,3}$, Kamran Kaveh$^1$, Mohammad Kohandel$^{1,2}$,} \\ {\sc Siv Sivaloganathan$^{1,2}$} \\ [2pt]  $^1$ Department of Applied Mathematics, University of Waterloo, \\ Waterloo, Ontario, N2L 3G1, Canada\\ [2pt] $^2$   Centre for Mathematical Medicine, Fields Institute, \\ Toronto, Ontario, M5T 3J1, Canada \\ [2pt] $^3$ School of Medicine, Queen's University, \\ Kingston, Ontario, Canada, K7L 3N6}
%\date{}
%\pagestyle{headings}
\markboth{A. DHAWAN ET AL.}{\rm A Stochastic Model for Tumor Control Probability }

%\titlerunning{Stochastic Model for Tumor Control Probability}
%\authorrunning{Andrew Dhawan et al}
\maketitle

\begin{abstract}
{Estimating the required dose in radiotherapy is of crucial importance since the administrated dose should be sufficient to eradicate the tumor and at the same time should inflict minimal damage on normal cells. The probability that a given dose and schedule of ionizing radiation eradicates all the tumor cells in a given tissue is called the tumor control probability (TCP), and is often used to compare various treatment strategies used in radiation therapy. In this paper, we aim to investigate the effects of including cell-cycle phase on the TCP by analyzing a stochastic model of a tumor comprised of actively dividing cells and quiescent cells with different radiation sensitivities. We derive an exact phase-diagram for the steady-state TCP of the model and show that at high, clinically-relevant doses of radiation, the distinction between active and quiescent tumor cells (i.e. accounting for cell-cycle effects) becomes of negligible importance in terms of its effect on the TCP curve. However, for very low doses of radiation, these proportions become significant determinants of the TCP. Moreover, we use a novel numerical approach based on the method of characteristics for partial differential equations, validated by the Gillespie algorithm, to compute the TCP as a function of time. We observe that our results differ from the results in the literature using similar existing models, even though similar parameters values are used, and the reasons for this are discussed.} %\linebreak \linebreak
{Radiotherapy, Tumor Control Probability, Cell Cycle, Mathematical Modeling, Stochastic Birth-Death Process, Method of Characteristics, Gillespie Algorithm}
\end{abstract}

\section{Introduction}
 External beam radiotherapy remains one of the most common treatment options for various cancers. However, the dose distribution of radiation must be optimized to reduce the risk of side effects of radiotoxicity and damage to healthy tissues surrounding the tumour volume. A widely used model for radiation treatment is the linear-quadratic (LQ) model \citep{LQ,munro}. This model estimates the surviving fraction of cancer cells after each treatment based on the total dose, and has the form:
\be
S(D) = e^{-\alpha D - \beta D^{2}},
\label{LQ}
\ee
where $\alpha$ and $\beta$ are sensitivity parameters (which depend on the tissue and the type of the applied beam) and $D$ is the total dose delivered during the radiation treatment. To include stochastic effects, a binomial or Poisson model has been used to describe the random variable representing the number of surviving cells after a treatment, centered upon a mean value determined by the linear-quadratic model of cell survival (see, for example, \cite{kallman,poisson}). An iterated birth and death process has been also suggested as a model of radiation cell survival \citep{Hanin2001}. A related quantity of interest is the tumor control probability (TCP) which is the extinction probability of the clonogenic cell population after radiation therapy. A model for the TCP accounting for cell proliferation dynamics was suggested by \citet{zaider}. Their model is a birth-death process for the probability distribution function of the tumor cells, $p_{n}(t)$, and the corresponding master equation of such a birth-death model is:
\be
\frac{{\rm d}p_{n}(t)}{{\rm d}t} = \lambda(n-1)p_{n-1}(t) + \zeta(n+1)p_{n+1}(t) - (\lambda + \zeta)np_{n}(t),
\label{zaider}
\ee
where $\lambda$ and $\zeta$ are the birth and death rates, respectively, and $n$ is the population of tumor cells. The effect of radiation is reflected as a time-dependent part in the death rate, $\zeta = \zeta_{0} + h(t)$, where $h(t)$ is known as the hazard function and is related to the radio-sensitivity parameters $\alpha$ and $\beta$ through the LQ model (Eq.\ref{LQ}). From Eq.\ref{zaider}, Zaider and Minerbo were able to calculate the extinction probability, $p_{0}(t)$, as a function of time and dose fractions (which is encoded in the form of $h(t)$). Thus, in their model, the TCP is given by:
\be
{\rm TCP}(t) = \left[1 - \displaystyle \frac{S(t)e^{(\lambda - \zeta)t}}{\displaystyle 1 + \lambda S(t) e^{(\lambda - \zeta)t}\int_{0}^{t}{\rm d}z (S(z)\exp{(\lambda - \zeta)z})^{-1}}\right]^{n_{0}}
\ee
where $n_{0}$ is the initial number of tumor cells and $S(t)$ is the exponential of the integral of the hazard function:
\bea
S(t) = \exp{ \int_{0}^{t} h(z){\rm d}z},\nonumber\\
h(D(t)) = (\alpha + 2\beta D)\frac{{\rm d}D}{{\rm d}t},
\eea
\nd with $D(t)$ being the dose in Gy delivered until time $t$ and its time derivative representing dose rate (Gy/day).\newline

\citet{hillen} expanded this approach to include the effect of cell cycle sensitivity in the TCP. They considered a two-compartment model for the active ($M, G_{1}, S$, and $G_{2}$ phases) and the quiescent ($G_{0}$ phase) cells (see also \citet{Gong2013}). The radio-sensitivity of resting cells and active cells are significantly different; the radio-sensitivity is typically much higher for actively proliferating cells \citep{leith}. This model was discussed both deterministically and stochastically in \citet{hillen}, but the stochastic master equation is solved under the assumption that the joint probability distribution function of two populations, $p_{n_{a},n_{q}}$, can be written in a factorized form as if the two random variables $n_{a}$ and $n_{q}$ are independent. However, this is clearly not true for small tumor populations, as pointed out by \citet{maler}. Small tumor populations can arise from a number of possible clinically relevant scenarios; for example, this would be the case for adjuvant radiation applied after surgery or chemotherapy, irradiation of micrometastases, as well as at the final stages of radiation therapy, when the tumor has shrunk to a few milimeters in size. Thus, as one approaches the limit of small tumor cell populations, a proper stochastic approach is needed to estimate the extinction probability, i.e. the TCP. Moreover, in previous cell cycle models of the TCP \citep{hillen,maler}, it is assumed that the proliferation is such that upon each cell division the daughter cells go into the $G_{0}$ (quiescent) state soon thereafter. In the following, we consider a more general situation where there is a probability $f$, such that one of the daughter cells goes into the resting phase upon division \citep{hillen2}; the master equation is again solved with the same assumption of independent random variables for the subpopulations of cells which breaks down in the key limit of small cell populations. In the following, we investigate thoroughly the TCP for such a model throughout the range of pertinent parameter values and plot a phase diagram of the model using a generating function method (see Sec. 2). In Sec. 3, we solve the differential equation for a probability generating function for the number of tumor cells using a novel final-value method of characteristics and in Sec. 4 we validate this with a Gillespie algorithm solution of the master equation.

\section{Stochastic two-compartment model with (a)symmetric proliferation}
Here we consider a two compartment model of active cells (A) and quiescent cells (Q), with the following dynamics: active cells can divide into either: (1) two quiescent cells or (2) one quiescent and one active, or (3) two active cells; assuming each active offspring is born with probability $f$ and each quiescent with probability $1-f$ while the proliferation rate for active cells is $\mu$. Note also that quiescent cells may, after a certain time, move from the $G_0$ to the $G_1$ phase of the cell cycle, and thereby become active. We assume this happens at a constant rate $\gamma$. Death rates for the cells in the active and quiescent compartments are denoted by $\Gamma_{a}$ and $\Gamma_{q}$, respectively:
\bea
A &\rightarrow& A+A:~~~~~~\mu f^{2}\nonumber\\
A &\rightarrow& A+Q:~~~~~~2\mu f(1-f)\nonumber\\
A &\rightarrow& Q+Q:~~~~~~\mu (1-f)^{2}\nonumber\\
Q &\rightarrow& A:~~~~~~~~~~~~~~\gamma \nonumber\\
A &\rightarrow& \o:~~~~~~~~~~~~~~~\Gamma_{a} \nonumber\\
Q &\rightarrow& \o:~~~~~~~~~~~~~~~\Gamma_{q}.
\label{model}
\eea
The deterministic ordinary differential equations (ODEs) for the above dynamics are given by:
\bea
\frac{{\rm d}n_{a}}{{\rm d}t} &=& -\mu f^{2} n_{a} + \gamma n_{q} - \Gamma_{a}(t) n_{a},\nonumber\\
\frac{{\rm d}n_{q}}{{\rm d}t} &=& 2(1-f)(1+f)\mu n_{a} -\gamma n_{q} -\Gamma_{q}(t)n_{q},
\label{deterministic}
\eea
where $n_{a,q}$ are the population of the active and quiescent compartments. The death rates of active and quiescent cells, $\Gamma_{a,q}$, are dose-dependent through the LQ formula (Eq.\ref{LQ}) and the given radiation protocol. Similarly, we can determine the stochastic dynamics of the model Eq.\ref{model} as follows. Denoting the joint probability distribution of having a population of $n_{a}$ active cells and $n_{q}$ of quiescent cells at time $t$ by $p_{n_{a}, n_{q}}(t)$, the master equation then reads,
\bea
\frac{{\rm d}p_{n_{a}, n_{q}}(t)}{{\rm d}t} &=& \mu f^{2}(n_{a}-1) p_{n_{a}-1,n_{q}}(t) + 2\mu f(1-f)n_{a}p_{n_{a},n_{q-1}}(t)\nonumber\\
&+&  \mu (1-f)^{2}(n_{a}+1) p_{n_{a}+1,n_{q}-2}(t) + \gamma(n_{q}+1) p_{n_{a}-1,n_{q}+1}(t)\nonumber\\
&+& \Gamma_{a}(n_{a}+1) p_{n_{a}+1,n_{q}}(t) + \Gamma_{q}(n_{q}+1) p_{n_{a},n_{q}+1}(t)\nonumber\\
&-& (\Gamma_{a} + \mu)n_{a} p_{n_{a},n_{q}}(t) - (\Gamma_{q} + \gamma)n_{q} p_{n_{a},n_{q}}(t).
\label{master}
\eea
The model in \citet{hillen} and \citet{maler} corresponds to $f=0$ in Eq.\ref{master}, while the Zaider and Minerbo model \citep{zaider} corresponds to $f =1$. We define the probability generating function for the joint probability distribution, $p_{n_{a},n_{q}}$,
\be
V(a,q,t) = \displaystyle \sum_{n_{a},n_{q}\geq 0} p_{n_{a},n_{q}}(t)a^{n_{a}}q^{n_{q}}.
\label{gf}
\ee
Using Eq. \ref{master} and Eq. \ref{gf}, we obtain the following partial differential equation (PDE) for $V(a,q,t)$:
\bea
\frac{\partial V}{\partial t} &=& \left[ \mu f^{2}\cdot a^{2} + 2\mu f(1-f)\cdot aq + \mu (1-f)^{2}\cdot q^{2}
-(\Gamma_{a} + \mu)a + \Gamma_{a}\right]\frac{\partial V}{\partial a}\nonumber\\
&+& \left[ \gamma\cdot a -(\Gamma_{q} + \gamma)q +\Gamma_{q}\right]\frac{\partial V}{\partial q}.
\label{pde}
\eea
Taking $n_{a,0}$ and $n_{q,0}$ to be the initial numbers of active and quiescent cells, respectively, we have the initial condition $V(a, q, 0) = a^{n_{a,0}}q^{n_{q,0}}$ and the boundary condition $V(1,1,t) = 1$, where the boundary condition comes from the definition of the generating function.\newline
\begin{figure}[h]
\begin{center}
\epsfig{figure=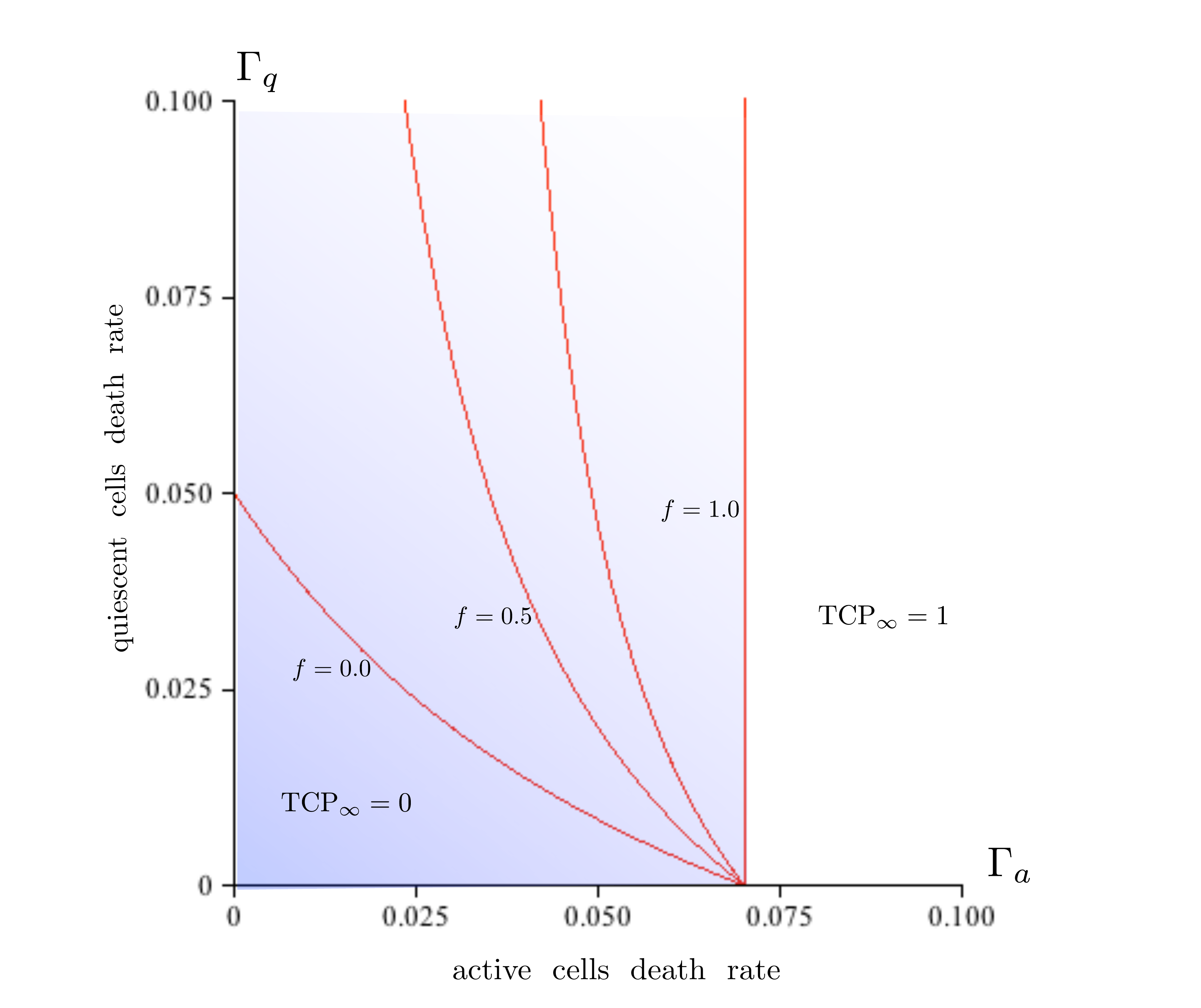, height=280pt,width=350pt,angle=0}
\end{center}
\caption{Phase boundaries for $\Gamma_{a}$ and $\Gamma_{q}$ with the active-cell division rate $\mu = 0.065 /{\rm day}$ and quiescent conversion rate $\gamma= 0.05 /{\rm day}$. Phase boundaries are plotted for various values of the asymmetric division factor, $f= 0.0, 0.5,0.7~{\rm and}~1.0$.}
\label{phase-gammaAgammaQ}
\end{figure}

\begin{figure}[h]
\begin{center}
\epsfig{figure=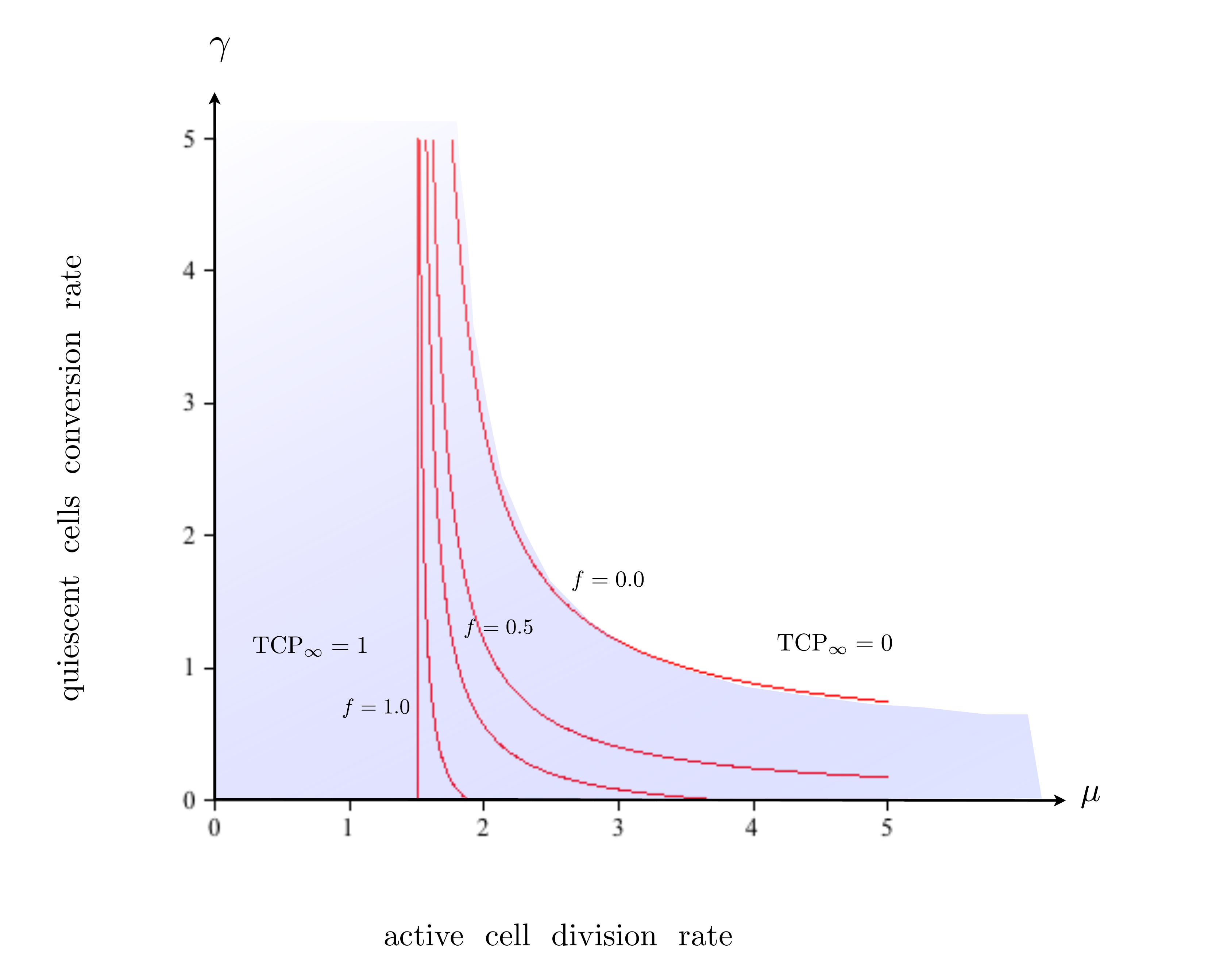, height=320pt,
width=370pt,angle=0}
\end{center}
\caption{Phase boundaries in the $\mu-\gamma$ plane. The death rates are fixed by the values used in \citet{hillen}. Phase diagrams are plotted for various values of asymmetric division factor, $f= 0.0, 0.5,0.7,0.9,1.0$.}
\label{phase-munu}
\end{figure}
In the case of a constant radiation dose, the TCP can be calculated in the steady state and we can find an analytical solution that relates TCP to all the parameters appearing in the model, especially the values of the death rates, $\Gamma_{a}$ and $\Gamma_{q}$. In the limit of a large - but finite - total number of cells $N$, we expect the steady state of the system to have two absorbing states of either zero population of either active or quiescent cells or both populations together reaching their maximum limits, $N_{a}$ and $N_{q}$ ($N_{a,q} \gg 1$). This means that in the steady state, the form of the generating function $V(a,q,t\rightarrow \infty)$ is:
\be
V(a^{*},q^{*}) = A + B\left(a^{*}\right)^{N_{a}}\left(q^{*}\right)^{N_{q}}.
\ee
The first term indicates that there is a non-zero probability for either population to become extinct and the second term is indicative of the possibility that eventually one or both populations reach large population limits - details to be determined by the values of $N_{a}$ and $N_{q}$. The coefficients $A$ and $B$ are the extinction and survival probabilities of the dynamical system, respectively, while $q^{*}$ and $a^{*}$ are the fixed points of Eq.\ref{pde}, which satisfy the following relations:
\bea
0 &=& \gamma a^{*} + \Gamma_{q} - (\Gamma_{q} + \gamma)q^{*}\nonumber\\
0 &=& \mu\left( fa^{*} + (1-f)q^{*}\right)^{2} - (\Gamma_{a} +\mu)a^{*} +\Gamma_{a}.
\label{fp}
\eea
The solutions for $a^{*}$ and $q^{*}$ are given by:
\bea
q^{*} &=& \displaystyle \frac{\gamma a^{*} + \Gamma_{q}}{\Gamma_{q} +\gamma},\nonumber\\
a^{*} &=& \displaystyle \frac{-c_{2} \pm \sqrt{c_{2}^{2} - 4c_{1}c_{3}}}{2c_{1}},
\eea
where the coefficients $c_1$, $c_2$, and $c_3$ are defined as
\bea
c_{1} &=& \displaystyle\frac{\mu}{(\Gamma_{q}+\gamma)^{2}}\left[ f\cdot(\Gamma_{q}+\gamma) + (1-f)\gamma\right]^{2},\nonumber\\
c_{2} &=& \displaystyle\frac{2\mu\Gamma_q(1-f)}{(\Gamma_{q} +\gamma)^{2}}\left[ f\cdot(\Gamma_{q}+\gamma) + (1-f)\gamma \right]-(\Gamma_{a} +\mu),\nonumber\\
c_{3}&=&\displaystyle \frac{\mu(1-f)^{2}\Gamma_{q}^{2}}{(\Gamma_{q}+\gamma)^{2}} + \Gamma_{a}.
\eea
Using the initial and boundary conditions mentioned above, we can obtain the values of $A$ and $B$. We are interested in the value of $A$, which is the extinction probability in the long run. This is the TCP in the steady state (${\rm TCP}_{\infty}$):
\be
{\rm TCP}_{\infty} = \left(a^{*}\right)^{n_{a},0}\left(q^{*}\right)^{n_{q},0}.
\label{tcp-infty}
\ee
The two fixed points of the system are $(1,1)$ and $(a^{*},q^{*})$. In parameter space, the phase boundary can be defined in the parameter space in terms of the model parameters such as $\Gamma_{a,q}$, $\gamma$, and $\mu$, when these parameters are such that $(a^{*}, q^{*}) = (1,1)$. For the region of the phase diagram where ${\rm TCP}_{\infty}=0$, the $(a^{*}, q^{*})$ fixed point is attractive while the $(1,1)$ fixed point is a saddle-point. As the parameters such as death rates $\Gamma_{a,q}$ increase, one moves into the ${\rm TCP}_{\infty}=1$ regime where now the fixed-point $(a^{*}, q^{*})$ vanishes and the only fixed point is $(1,1)$ which is globally attractive. The phase boundary for variable death rates is plotted in Fig.\ref{phase-gammaAgammaQ}. To provide a comparison between the results of \citet{hillen}, we use identical parameter values, namely a constant radiation dose rate of $R(t)=2.75$ Gy/day, and the division rate $\mu$  and the conversion rate $\gamma$ were taken to be $0.065~{\rm day}^{-1}$ \citep{swanson} and $0.047~{\rm day}^{-1}$ \citep{basse}, respectively. Death rates, which were effectively derived from a limit of the LQ model, are given by: $\Gamma_{q} = 0.4/{\rm Day}$ and $\Gamma_{a} = 1.5/{\rm Day}$. The death rates were derived by using the dose-dependent survival fraction given by the LQ model, creating a hazard function from that, and substituting in values for the radiosensitivity parameters $\alpha_a = 0.487 \rm Gy^{-1}$, $\alpha_q = 0.155 \rm Gy^{-1}$, $\beta_a = \beta_q = 0.055 \rm Gy^{-2}$ taken from \citet{leith}, where the subscript $a$ or $q$ indicates active cells or quiescent cells, respsectively. Also, note that a constant radiation dose is not necessarily a clinical possibility for treatment, but is used in order to facilitate direct comparison with the results of \citet{hillen}.\newline

\nd Our plots in Fig.\ref{phase-gammaAgammaQ} for the phase boundary between ${\rm TCP = 0}$ and ${\rm TCP = 1.0}$ regimes show the interesting evolution of the two regimes of the one-compartment model of \citet{zaider}
 into the two-compartment model of \citet{hillen}. It can be noted that the two ends of the phase boundary at the $\Gamma_{a}$-axis and $\Gamma_{q}$-axis are in fact $\mu$ and $\gamma$ for the fully two-compartment model ($f=0$), i.e. the values for the cutoff death rates are determined by the proliferation and conversion potentials $\mu$ and $\gamma$.For values of $\Gamma_{a,q}$'s in these regions one expects to get an unsuccessful therapy or ${\rm TCP}_{\infty}=0$. The implication of this is the fact that the values of $\Gamma_{a,q}$ estimated for real irradiation protocols lie deep inside the ${\rm TCP}_{\infty} = 1$ phase for all the values of the asymmetric proliferation factor, $f$, and thus the division of the population into different compartments based on the cell-cycle has a 
negligible effect on the TCP, given that the single and two compartment models utilize identical parameters. That is, given a real treatment schedule, the effect of $f$ on the TCP curve itself becomes negligible. \newline

\nd We have also plotted the phase boundary for ${\rm TCP}_{\infty} = 0,1$ for different values of the division and conversion rates $\mu$ and $\gamma$ in Fig.\ref{phase-munu}. A similar evolution between a one-compartment and two-compartment model can be observed in this case. The phase boundaries for the ${\rm TCP = 0}$ and ${\rm TCP = 1}$ regimes can be used to determine a crude cutoff dose below which treatments will not work, and above which treatments will work in finite time. However, we note that for clinical treatments, parameter values must be deep inside the ${\rm TCP = 1}$ regime to succeed within a reasonable timescale. In the next two sections we will focus on the time-dependence of the TCP via two different approaches.\newline

\section{Numerical solutions: Final-value method}
In the previous section, we discussed the steady-state behavior and the fixed points of Eq.\ref{pde}. In this section, we derive the time dependence of the TCP as it approaches unity for a given radiation protocol. Solving (Eq.9), i.e. the PDE for the generating function, with a combination of initial and boundary conditions is a difficult task. We approach the problem by a novel application of the method of characteristics. Consider a PDE of the form:
\begin{eqnarray}
\frac{dV}{dt} = \frac{\partial V}{\partial x_1} f_{x_1}(x_1,+ \cdots + x_n,t)+ \cdots +
\frac{\partial V}{\partial x_n} f_{x_n}(x_1,+ \cdots + x_n,t).
\end{eqnarray}
Recall that the method of characteristics relies upon finding a set of characteristic curves $t(s),x_1(s),\cdots,x_n(s)$ such that $f(s)=V(x_1(s),x_2(s),\cdots,x_n(s),t(s))$ is a constant. Then, by the chain rule:
\begin{eqnarray}
\frac{{\rm  d}f}{{\rm  d}s} = \frac{\partial V}{\partial x_1} \frac{{\rm  d} x_1}{{\rm  d} s} + \frac{\partial V}{\partial x_2} \frac{{\rm  d} x_2}{{\rm  d} s} + \cdots + \frac{\partial V}{\partial x_n} \frac{{\rm  d} x_n}{{\rm  d} s} + \frac{\partial V}{\partial t} \frac{{\rm  d} t}{{\rm  d} s} = 0 .
\end{eqnarray}
By comparing the form of this differential equation with the form of the equation we wish to solve, we observe that to find these characteristic curves, the following set of ordinary differential equations must be solved:
\begin{eqnarray*}
\frac{{\rm  d}x_1}{{\rm  d} s} &=& f_{x_1}(x_1(s),x_2(s), \cdots, x_n(s),t(s))\\
\frac{{\rm  d} x_2}{{\rm  d} s} &= &f_{x_2}(x_1(s),x_2(s), \cdots, x_n(s),t(s))\\
\vdots& \\
\frac{{\rm  d} x_n}{{\rm  d} s} &=& f_{x_n}(x_1(s),x_2(s), \cdots, x_n(s),t(s))\\
\frac{{\rm  d} t}{{\rm  d} s} &=&-1 .
\end{eqnarray*}
Note that we constrain $t(0)=0$, so that the initial conditions of the system can be used in the calculation of $f(0)$. The last equation in the system, given the initial condition $t(0) =0$, can be solved. Thus we obtain the following system:
\begin{eqnarray}
\frac{{\rm  d} x_1}{{\rm  d} t} &=& f_{x_1}(x_1(t),x_2(t), \cdots, x_n(t),t) \nonumber \\
\frac{{\rm  d} x_2}{{\rm  d} t} &= &f_{x_2}(x_1(t),x_2(t), \cdots, x_n(t),t) \nonumber \\
\vdots&  \nonumber \\
\frac{{\rm  d} x_n}{{\rm  d} t} &=& f_{x_n}(x_1(t),x_2(t), \cdots, x_n(t),t) .
\label{eqn:ODEs}
\end{eqnarray}
We also notice that for this particular set of characteristic curves,
\begin{eqnarray}
f(s)=V(x_1(s),x_2(s),\cdots,x_n(s),t(s)) = f(0) = V(x_1(0),x_2(0), \cdots, x_n(0),0).
\end{eqnarray}
We define $f_0$ as the function relating the initial values of the characteristic functions to the initial conditions for the PDE. From the given initial condition for our PDE, we have
\begin{eqnarray}
V(x_1(0),x_2(0), \cdots, x_n(0),0) = f_0(x_1(0),x_2(0),\cdots,x_n(0)).
\end{eqnarray}
This gives $f(s) =  f_0(x_1(0),x_2(0),\cdots,x_n(0))$.

Recall that we are only interested in the function $g(t) = V(0, \cdots, 0, t)$, and not the entire solution to the PDE since $g(t)$ represents the extinction probability of the tumor at the time $t$, which is exactly the TCP. Thus, to compute $g$ at a fixed $t = t^*$, the only characteristic curve that needs to be considered is such as $x_1(t^*) = x_2(t^*) = \cdots = x_n(t^*) = 0$. We denote these characteristic curves $\bar{x}_1, \bar{x}_2, \cdots, \bar{x}_n$. Moreover, based on the above discussion, we observe that
\begin{eqnarray}
g(t^*)=f_0(\bar{x}_1(0),\bar{x}_2(0),\cdots,\bar{x}_n(0)).
\end{eqnarray}
The values $\bar{x}_i(0)$ are determined by the set of ODEs in (\ref{eqn:ODEs}), with the final value condition that $x_1(t^*) = x_2(t^*) = \cdots = x_n(t^*) = 0$. Thus, to obtain $g(t)$, at any set of time points, the final value problem must be solved independently to obtain the initial values of the characteristic curve, which must then be substituted into the initial condition for the PDE.

We note that taking $t\rightarrow t^*-t$ will transform the aforementioned final value problem into an initial value problem, where the desired values become $\bar{x}_i(t^*)$. In this case, notice that the computation of the function $g(t)$ can be \emph{vastly} simplified if the functions $f_i(x_1(t),x_2(t),\cdots,x_n(t),t)$ do not depend explicitly on $t$. That is, if $f_i(x_1(t),x_2(t),\cdots,x_n(t),t) = \hat{f}_i(x_1,x_2,\cdots,x_n)$, then observe that for every $t^*$, the set of ODEs that must be solved is the same, and all have the same initial condition that $\bar{x}_i=0$. Thus, in this case, computation of the function $g(t)$ can be done for all $t$ in a given interval, by solving the set of coupled ODEs \emph{once}. If this simplification cannot be made, then the method will still solve the PDE, but for each time point, the set of ODEs that must be solved will be different.

\section{Gillespie solution}
In order to simulate the stochastic process representing the cellular dynamics within the model framework, Gillespie's algorithm for stochastic simulation was implemented. This algorithm simulates one realization of the time evolution of the system by first computing propensities for the events that can occur at any time step (i.e. the set of cell births/deaths in the above model). Subsequently, the time before the next event occurs is computed via an exponential distribution, and the event that occurs at this time step is chosen by a distribution weighted by the total propensity of all events (i.e. the likelihood that any reaction would occur). Thus, the events occur individually, with a likelihood proportional to their individual propensity, and the times between the individual events is based on an exponential distribution of waiting times, weighted by the total propensity of all events. Each simulation describes one specific time course for the system. This is then repeated a large number of times, typically $10^5$ in our simulations, and for each, an indicator function known as the treatment success indicator is defined: $\rm{TS}_i (t)=1$ if at time $t$, the tumor is controlled (i.e. there are zero cells remaining), and $0$ otherwise. Then, after $M$ such simulations, the TCP function is defined to be:
\begin{equation}
\rm{TCP}(t)=\frac{1}{M}\displaystyle \sum_{i=1}^M \rm{TS}_i(t) .
\end{equation}
The process to calculate $\rm{TS}_i(t)$ is: (1) Compute likelihood of each cellular reaction occurring ($L_i$ for reaction $R_i$). (2) Sum together all likelihoods into quantity $T_L=\sum_i L_i$. (3) Compute uniformly distributed random numbers $p_1$ and $p_2$ in the interval $(0,1)$. (4) Compute the next time step of a likelihood reaction, assuming exponentially distributed times $dT=-\ln(p_1)/T_L$. (5) Update time variable by adding time step computed $t=t+dT$. (6) Determine which reaction to carry out: if $L_{i-1}/T_L\leq p_1 \leq L_i/T_L$, carry out reaction $R_i$. (7) Update the cellular population variables, assuming reaction $R_i$ was carried out. (8) If number of stem cells is zero, treatment success is one and terminate program, else treatment success is zero and repeat step $1$. (9) If time  is greater than the max simulation time, treatment success is zero and terminate program.

We illustrate the effectiveness of the numerical method presented in solving for the TCP for the active quiescent model that was outlined previously. To do this, we compare the TCP as computed by a high number of Gillespie simulation runs with the TCP as computed by the output of the numerical method.

\begin{figure}
\includegraphics[scale=0.5]{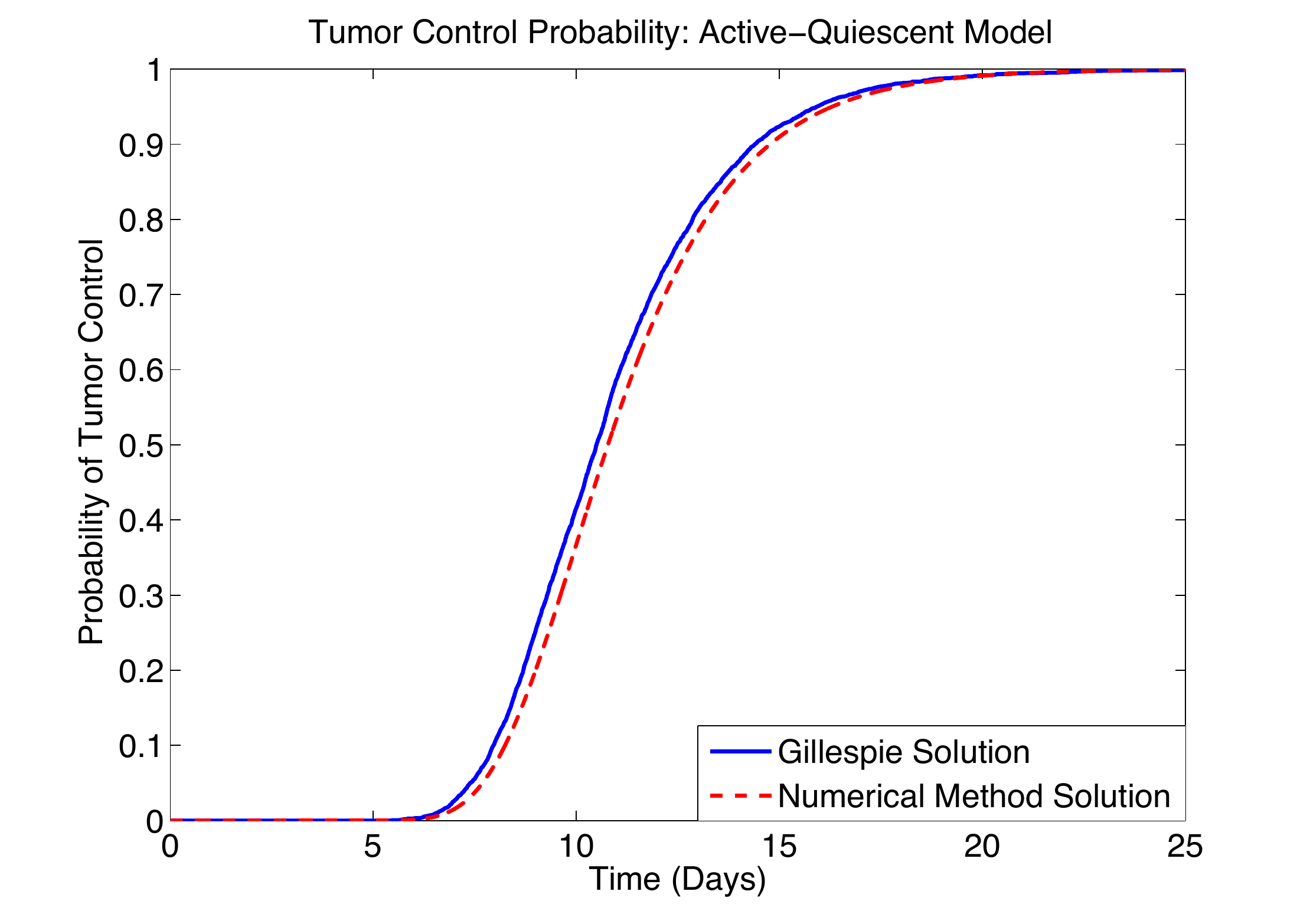}
\caption{A plot of the TCP computed by the numerical method outlined above, and by Gillepsie's algorithm.}
\label{fig:TCPplot}
\end{figure}

To obtain a proper stochastic limit, we use a small number of each type of cell, letting $a_0 = 10^2 = q_0$. Using these and the rest of the parameter values mentioned in Sec. 2, we obtain the TCP plot depicted in Fig.~\ref{fig:TCPplot}. In this plot, both the numerical solution, computed by an implementation of the method presented above, as well as the Gillespie solution are plotted, to highlight the high degree of similarity between the curves. In order to quantify the degree to which these curves agree, we sample both curves at the nine time points corresponding to $t=0, 3, 6, \cdot, 24$ and compute a root-mean-square distance between the two vectors representing the TCP values of the Gillespie and numerical solutions to obtain $0.022$, which is indeed very small.

\begin{figure}[!]
\includegraphics[scale=0.38]{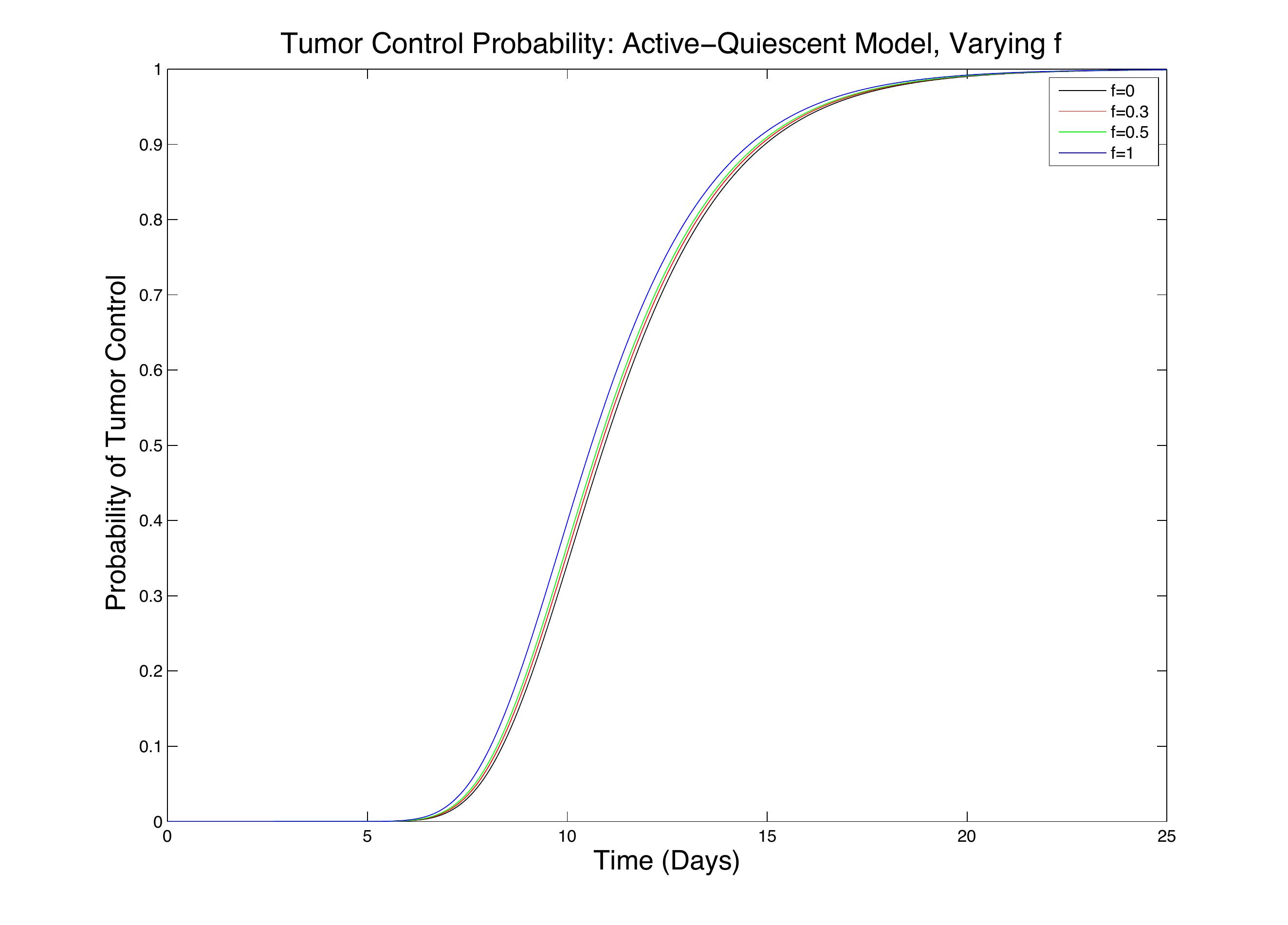}
\caption{A plot of TCP vs. time for different values of asymmetric division factor, $f$. As was predicted in Sec. 2, all the graphs coincide. The parameter values are from \citet{hillen}. The value for the dose delivery rate, $(R(t) = 2.75 {\rm Gy}/{\rm day})$, is so high that the differences between different TCP plots are indistinguishable. }
\label{fig:TCPplot2}
\includegraphics[scale=0.51]{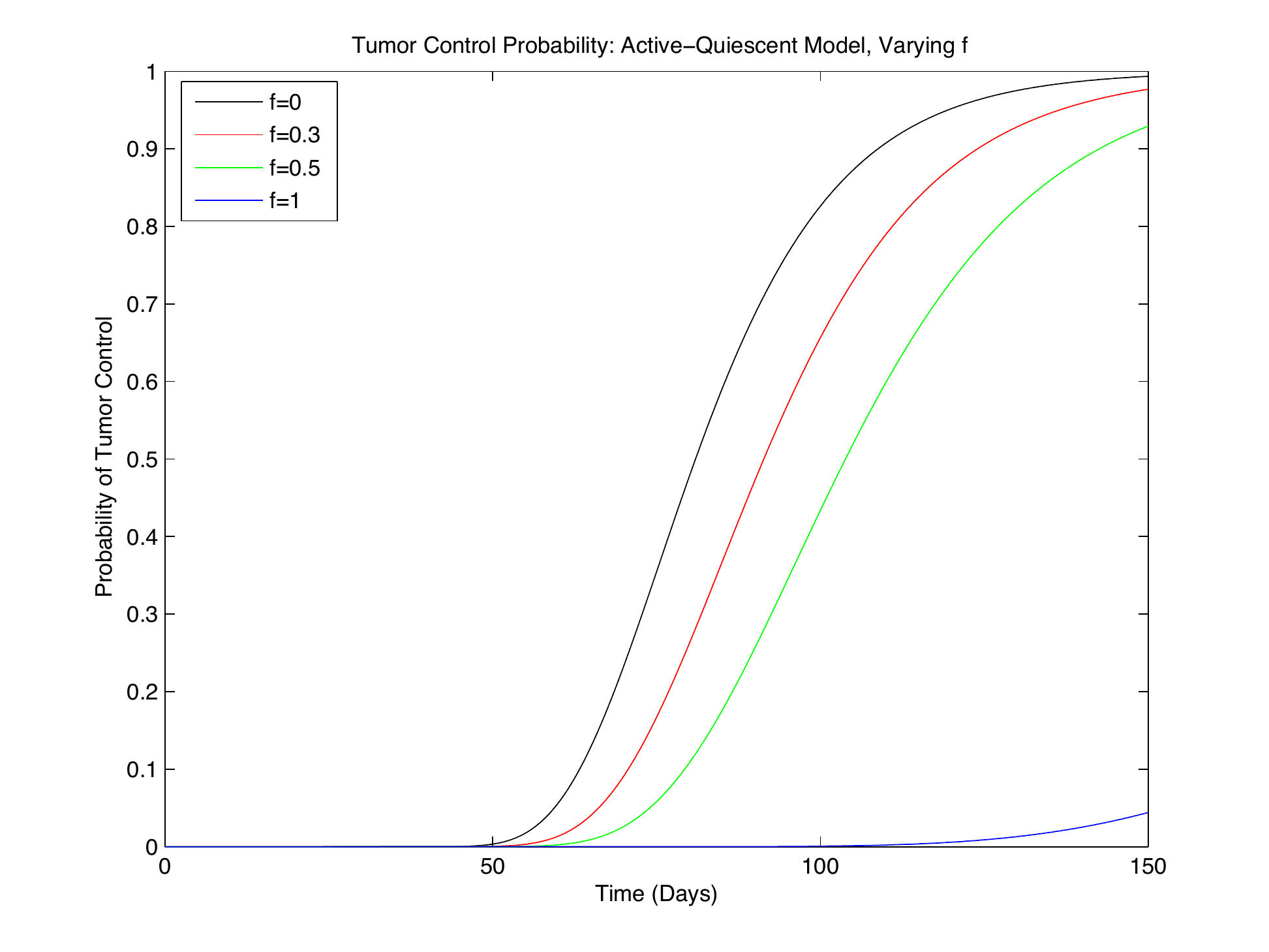}
\caption{A plot of TCP vs. time for different values of asymmetric division factor, $f$. The values for death rates are chosen to be $\Gamma_{a}= 0.08$ and $\Gamma_{q}=0.1$. These values give distinct TCP graphs and particularly ${\rm TCP}_{\infty}(f=1)$ almost not reaching unity.}
\label{fig:TCPplot3}
\end{figure}

Next, to check the relevance of the two-compartment model, we plot the TCP vs. time for different values of asymmetric division, $f$. As discussed in Sec. 2, we do not expect any difference as the physical parameters estimated from clinical data indicate a high-death rate for both the active and quiescent cells which lie deep inside the overlap region of the one-compartment and two-compartment models. As shown in Fig.\ref{fig:TCPplot2}, this is in fact the case and the ${\rm TCP}(t; f)$ plots are almost indistinguishable.

However, if one decreases both death rates, from the values in the phase diagram in Fig.\ref{phase-gammaAgammaQ}, one should expect any difference between the ${\rm TCP}(t; f)$ to reveal itself. One example is plotted in Fig.\ref{fig:TCPplot3}, with death rates $\Gamma_{a} = 0.08/{\rm Gy}$ and $\Gamma_{a} = 0.1/{\rm Gy}$. Using the phase diagram, we can see that these values correspond to a point in the $\Gamma_{a} - \Gamma_{q}$ plane very close to the $f=1$ phase-boundary. This explains why the TCP graph for $f=1$ in Fig.\ref{phase-gammaAgammaQ} appears to approach unity on a much longer time scale than the other graphs. Similarly, one can expect the characteristic saturation time of the TCP (i.e. the time to reach unity) to tend to infinity as we choose death rates (by varying the dose of radiation) that cross the phase boundary corresponding to that asymmetric proliferation factor $f$.

\section{Discussion}
In this work, we have investigated a two-compartment stochastic model for the tumor control probability by including the asymmetric nature of division of active cells into either quiescent cells or active cells. We argue that the method suggested by \citet{hillen2} does not properly address the coupled nature of the joint probability distribution of the active and quiescent populations and have presented an alternative consistent approach. We have analytically derived all regimes of the phase diagram of ${\rm TCP}=0, 1$ in steady-state, for variable division and conversion rates and also separately the phase diagram of ${\rm TCP}=0, 1$ for varying death rates. From the phase diagram, we may conclude that the two-compartment model diminishes the effects of the original birth-death model of \citet{zaider} while the significantly lower death rates (dose delivery rates and radio-sensitivities) can be addressed with a two-compartment model which includes cell cycle effects. The phase boundaries obtained for the ${\rm TCP}=0$ and ${\rm TCP}=1$ regimes can be used to crudely determine a dose cutoff suitable for tumor control for tumors comprised of different populations of active and quiescent cells, when death rates are low enough between treatments being compared so that parameters such as $f$ become significant. We note that the time to achieve tumor control depends on the distance from the phase boundary, and those parameters within the ${\rm TCP}=1$ regime but very close to the phase boundary may not be able to achieve tumor control in a realistic time frame.

We also note now that there is a significant difference in the results computed via the method presented here and the results presented in \citet{maler} using similar parameter values. In \citet{maler}, the computed TCP curve shows that the time to cure for a population of 1000 cells in total is approximately 20 hours, which is much less than the 20 days predicted by the model presented here (for a smaller population of 100 cells).

To complete the analysis we have presented a comprehensive numerical approach to compute the TCP as a function of time. The numerical method (which we call the Final Value Method), when implemented to solve the TCP problem for the above case and parameter set, can be seen to solve the PDE, producing nearly identical solutions to that of the Gillespie algorithm, which is a good approximation to the true solution. Based on the work presented here, we may conclude that the final value method is a new way to numerically solve any PDE with an initial condition that is of a form appropriate for the method of characteristics. In the case presented above, this method has been utilized to solve the real-world problem of computing the TCP for a model based on incorporating cell-cycle effects into radiotherapy treatment planning, by using a two-compartment model for the active and quiescent cells.

One should note that the death rates described in this paper are dose-dependent death rates for radiotherapy, but could easily be interpreted as death rates from chemotherapy for instance. In fact, it is well-known that the cytotoxic effects of chemotherapy primarily impact cells actively proliferating within the cell cycle, so here the division between active and quiescent cell populations become important. Thus, one may anticipate that the framework presented in this paper can be extended to study the effects of other treatments for tumor control, such as chemotherapy.

\section*{Acknowledgements}
This work was financially supported by the NSERC/CIHR Collaborative Health Research Grant (to MK and SS).
%\end{Acknowledgements}
%\section{References}

%\bibliographystyle{/JMathBio/spbasic.bst}
%\bibliographystyle{/JMathBio/spmpsci.bst}

%\bibliographystyle{/JMathBio/spphys.bst}
\bibliographystyle{apalike}

\bibliography{DhawanTCP_MMB}

\end{document}